\documentclass[%
reprint,
superscriptaddress,
amsmath,amssymb,aps,
]{revtex4-1}

\usepackage{float}
\usepackage{graphicx}
\usepackage{dcolumn}
\usepackage{bm}
\usepackage{mathrsfs}
\usepackage{natbib}

\usepackage{color}

\newcommand{\fig}{FIG. }
\newcommand{\eqn}{eqn. }

\begin{document}


\title{Dynamics of nanoscale bubbles growing in a tapered conduit}

\author{Michael M. Norton}
 \email{mike.m.norton@gmail.com}
 \affiliation{University of Pennsylvania, Philadelphia, Pennsylvania 19104, USA}
 \affiliation{Brandeis University, Waltham, Massachusetts 02453}
\author{Nicholas M. Schneider}
 \affiliation{University of Pennsylvania, Philadelphia, Pennsylvania 19104, USA}
\author{Frances M. Ross}
 \affiliation{IBM T.J. Watson Research Center, Yorktown Heights, NY 10598, USA}
\author{Haim H. Bau}
 \email{bau@seas.upenn.edu}
 \affiliation{University of Pennsylvania, Philadelphia, Pennsylvania 19104, USA}

\date{\today}

\begin{abstract}
We predict the dynamics and shapes of nanobubbles growing in a supersaturated solution confined within a tapered, Hele-Shaw device with a small opening angle $\Phi \ll 1$. Our study is inspired by experimental observations of the growth and translation of nanoscale bubbles, ranging in diameter from tens to hundreds of nanometers, carried out with liquid-cell transmission electron microscopy. In our experiments, the electron beam plays a dual role: it supersaturates the solution with gaseous radiolysis products, which lead to bubble nucleation and growth, and it provides a means to image the bubbles in-situ with nanoscale resolution. To understand our experimental data, we propose a migration mechanism, based on Blake-Haynes theory, which is applicable in the asymptotic limits of zero capillary and Bond numbers and high confinement. Consistent with experimental data, our model predicts that in the presence of confinement, growth rates are orders of magnitude slower compared to a bubble growing in the bulk and that the combination of a tapered channel and contact line pinning create tear-drop shaped bubbles. 
\end{abstract}

\pacs{Valid PACS appear here}
\maketitle


\section{Introduction}

\begin{figure}
\includegraphics[width=1.0\columnwidth]{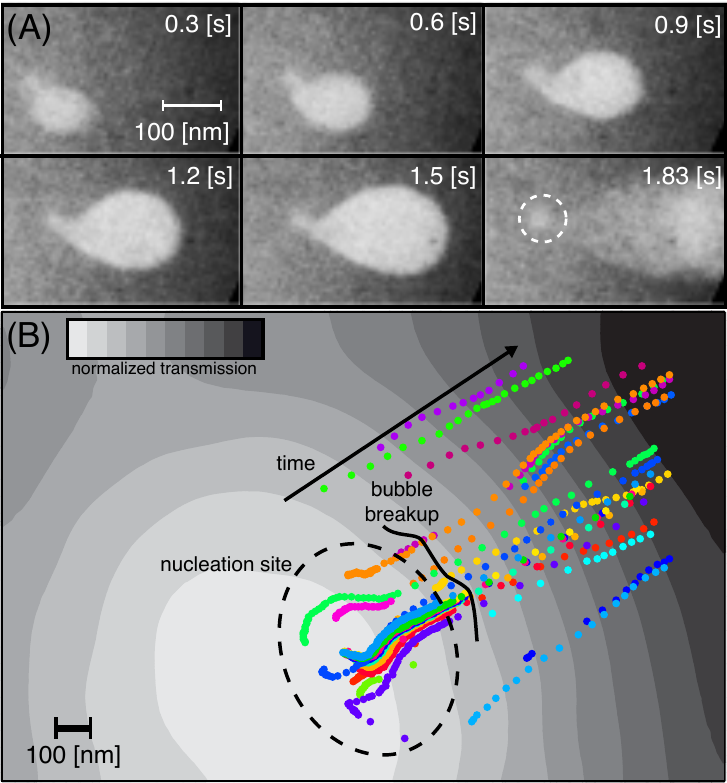}
\caption{(A) A series of transmission electron microscopy images of a bubble growing under high, non-uniform confinement witinh a liquid cell, \fig S1 and movie S1. The bubble formation occurred while imaging an aqueous solution of gold nanorods that contained a trace amount of the surfactant cetrimonium bromide (CTAB) with TEM at 300 keV, beam current $\sim$1−10 nA, and beam radius $\sim$2 $\mu$m. The bubble grows anisotropically and in the last frame detaches from the nucleation size (dashed white circle), the growth of this bubble is compared to theory in \fig\ref{fig:compareshapes}. (B) Normalized intensity (contours) with bubble trajectories (data points), nucleation site (dashed ellipse), and approximate location of bubble breakup (solid line).}\label{fig:growthsequence}
\end{figure}

The motion and shape of droplets and bubbles in confined geometries have attracted considerable attention in the scientific community due to their importance in industrial processes, multi-phase flows through porous media and recently, microfluidic devices. Typically at the macroscale, pressure driven flow or buoyancy drive motion of bubbles and fluid-fluid interfaces. G. I. Taylor and Saffman \cite{Saffman1958,Taylor1959} and Tanveer and Saffman \cite{Tanveer1987} studied the migration of droplets in Hele-Shaw cells. Matched asymptotic methods have subsequently been used to elucidate the geometry of gas bubbles in cylindrical tubes \cite{Bretherton1961} and in Hele-Shaw cells when the continuous phase either perfectly wets the surface \cite{Park1984} or when contact line dynamics are at play \cite{Weinstein1990}.

In the absence of applied pressure gradients and when drops are small such that the Bond number is much less than unity, gradients in substrate elasticity \cite{Style2012,Style2013}, chemistry \cite{Chaudhury1992, Darhuber2005}, temperature, and electric field, among others, can spontaneously drive the motion of contact lines and interfaces. Similarly, geometric gradients can induce capillary forces and promote transport of drops/bubbles confined in tapered capillaries \cite{Renvoise2009}  and residing on wires with varying cross-sections  \cite{Hanumanthu2006, Lorenceau2004}. To minimize their surface energy, wetting drops seek confinement while non-wetting drops and bubbles avoid it. The dynamics of bubbles and droplets in tapered Hele-Shaw devices have been modeled and compared to experiment \cite{Jenson2014,  Metz2009, Reyssat2014}. In these cases a dynamically created film wets the substrate. Furthermore, the mass of the bubble or drop is fixed and so the process is purely relaxational. Droplets and bubbles begin in a non-equilibrium configuration and transport to their equilibrium positions, where they can exist as spheres (for completely wetting substrates) or spherical sections (for partially wetting substrates) that satisfy the equilibrium contact angle \cite{Langbein2002}. Laplace pressure gradients drive the disperse phase and hydrodynamics either in the bulk or at the contact line provide dissipation, setting the time scale for the process.

In recent years, there has been a considerable interest in surface-bound nanobubbles and their stability  \cite{Alheshibri2016, Ball2012, Craig2011, Fang2016}.  Understanding such bubbles has the potential to improve processes such as acoustic surface scrubbing \cite{Liu2008, Liu2009, Yang2011}  and designing surface textures that optimize onset of nucleate boiling to enhance heat transfer \cite{Dong2014,Fazeli2015,Chu2012}. Free nano-bubbles can also be used as contrast agents in ultra-sonic imaging \cite{Cai2015} and so developing processes which reliably produce nanobubles could have applications in medical imaging as well. Due to the extremely high Laplace pressure of free bubbles, mass transfer-driven dissolution can dominate dynamics when the bubbles are below their critical radius. As the radius of curvature increases $R^{-1}$, the internal pressure $P$ and the equilibrium surface concentration $C_\text{s}$ rise; the latter in accordance with Henry's law,  $C_\text{s}\propto P\propto R^{-1}$.  This positive feedback results in rapid dissolution; a bubble 100nm in radius will dissolve in $\sim$100$\mu$s \cite{Epstein1950, Ljunggren1997}. In contrast to the above prediction, surface-bound nanobubbles have been observed to persist for many hours \cite{Zhang2008, Seddon2011, Sun2016}. Researchers have proposed various mechanisms for this anomalous behavior, including a perpetual dynamic equilibrium \cite{Petsev2013,Seddon2011}, the stabilizing effect of organic contaminants \cite{Zhang2012}, and contact line pinning \cite{Weijs2013,Lohse2015,Maheshwari2016}. The constraint of a pinned contact line introduces a negative feedback by forcing the radius of curvature to decrease as the mass of the bubble decreases. As the surface concentration of dissolved gases decreases and approaches that of the bulk concentration, mass transfer slows.


Inspired by this simple feedback mechanism between the geometry of a bubble and its \emph{growth}, we reaxmine the transport of bubbles in tapered Hele-Shaw cells when mass transfer and contact line dynamics dominate the system behavior. We develop our model around experimental observations of nanobubbles $R\sim 10-100$nm growing and migrating in tapered Hele-Shaw cells with plate gaps $\sim$10-100s of nanometers, \fig\ref{fig:growthsequence}\cite{Grogan2014}. The bubbles grow anomalously slow when compared to classical mass transfer limited growth theories \cite{Epstein1950} and despite their small size and velocities (the capillary numbers are in the range $10^{-9}-10^{-7}$ and Bond numbers $10^{-11}-10^{-9}$) are rarely spherical in shape. To explain both observations, we posit that the contact line is slow to relax compared to the rate of mass transfer and dominates bubble shape, growth rate, and translation in a regime where the gas-liquid interface is always at mechanical equilibrium (uniform Laplace pressure). We utilize the Blake-Haynes (BH) mechanism to model contact line movement. When a portion of the contact line is immobilized, a teardrop shaped bubbles result, similar to those observed in our experiments.

\section{Bubble Growth and Migration Model}
\begin{figure}[h]
\includegraphics[width=\columnwidth]{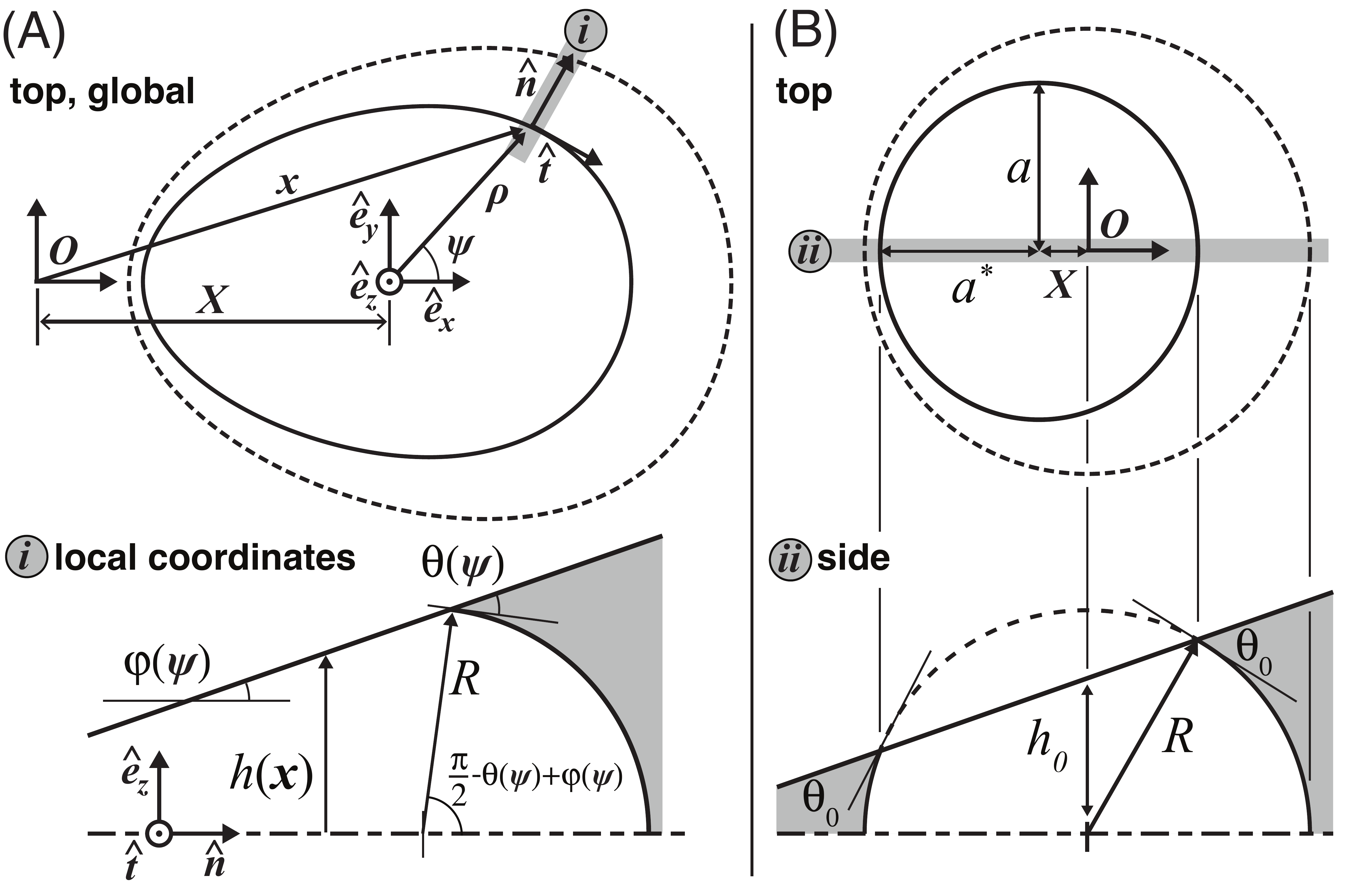}
\caption{: (A) Schematic of the geometry of our problem with global perspectives of a bubble shape (top) as seen from the top and a cross section of the interface as viewed from the local coördinate system (\emph{i}); the unit vectors of the latter are defined by the local tangent and normal vectors of the bubble's interface at the $xy$-plane; both the slope $\varphi$ and contact angle $\theta$ are local properties and depend on the polar angle $\psi$.  (B) Top and cross-section (through the plane \emph{ii}) view of the initial condition used for the model, a sphere with a circular contact line that when projected onto the $xy$-plane is an ellipse with major and minor radii $\{a,a^*\}$.}\label{fig:schematic}
\end{figure}

Our goal is to model contact line evolution of a bubble growing in a supersaturated solution confined between two plates diverging at an $2\Phi$.  \fig\ref{fig:schematic}A  depicts top view of the bubble. The bubble is symmetric with respect to the $x$-axis. The inner contour (solid line) is the projection of the bubble's contact line with the confining plates onto the $xy$-plane  and is defined with the polar coordinate $\rho\left(\psi,t\right)$, where $\psi$ is the azimuthal angle and $t$ is time. The intersection of the bubble's surface with the $xy$-plane is shown as the outer, dashed contour. \fig\ref{fig:schematic}B depicts the cross-section of the bubble confined in the tapered conduit. The two plates are distance $2h\left(\mathbf{x},t\right)$ apart. We restrict our analysis to cases when $R\left(t \right) \ll \min\limits_{\psi}\rho\left(\psi,t\right)$.  In the limit of zero capillary and Bond numbers, the pressure inside the bubble is nearly uniform and dominated by the smallest radius of curvature $R$. Hence, we take $R$ to be $\psi$-independent. $\theta\left(\psi\right)$ is the dynamic contact angle between the continuous phase (liquid) and the plates. The contact angle may vary as a function of position.

The projection of a contact line point on the $x-y$ plane is given by the two-dimensional vector
\begin{equation}
{\bf{x}}\left( {\psi ,t} \right) = \left\{ {\rho \left( {\psi ,t} \right)\cos \psi  + X\left( t \right),\rho \left( {\psi ,t} \right)\sin \psi } \right\}.
\label{eq:contactline}
\end{equation}
In the above, $X$ is the distance of the origin of $\rho$ from the center of the initial $\left(t=0\right)$ bubble geometry, which we will later define to be a sphere (\eqn\ref{eq:RIC}). $2h_0$ is the height of the conduit at $X=0$.

We define local planar coordinates aligned with the normal $\hat{\mathbf{n}}$ and tangent $\hat{\mathbf{t}}$ to the projection of the contact line on the $xy$-plane. Using \fig\ref{fig:schematic}\emph{i} as a guide, we relate the radius of curvature $R$ to the slope of the wedge, contact angle, and channel height at the contact line in the plane that is both normal to the contact line at position $\mathbf{x}$ and the $xy$- plane. 
\begin{equation}
h = R\cos \left( {\theta \left( \psi  \right) - \varphi \left( \psi  \right)} \right).
\label{eq:localheight1}
\end{equation}
Additionally,
\begin{equation}
h = {h_0} + x\tan \Phi  = {h_0} + \left( {X + \rho \cos \psi } \right)\tan \Phi.
\label{eq:localheight2}
\end{equation}
Equating \eqn\ref{eq:localheight1} and \ref{eq:localheight2} and solving for the contact angle $\theta$ gives
\begin{equation}
\theta  = {\cos ^{ - 1}}\left[ {\frac{{{h_0} + \left( {X + \rho \cos \psi } \right)\tan \Phi }}{R}} \right] + \varphi.
\label{eq:localcontactanlge}
\end{equation}
The local wedge angle $\varphi\left(\theta\right)$ relates to the global or maximum wedge angle $\Phi$ through
\begin{equation}
\tan \varphi  = {\bf{\hat n}} \cdot {{\bf{\hat e}}_x}\tan \Phi 
\label{eq:localslopeA}
\end{equation}
When  ${\left. {{\bf{\hat n}}} \right|_{\psi  = 0}} = {{\bf{\hat e}}_x}$ \eqn\ref{eq:localslopeA} yields ${\left. \varphi  \right|_{\psi  = 0}} = \Phi $; when ${\left. {{\bf{\hat n}}} \right|_{\psi  = \pi /2}} = {{\bf{\hat e}}_y}$ , ${\left. \varphi  \right|_{\psi  = \pi /2}} = 0$.  In radial coordinates, the normal vector to the contact line curve 
\begin{equation}
{\bf{\hat n}} = \left\{ {\begin{array}{*{20}{c}}
{\rho \cos \psi  + \frac{{\partial \rho }}{{\partial \psi }}\sin \psi }\\
{\rho \sin \psi  - \frac{{\partial \rho }}{{\partial \psi }}\cos \psi }
\end{array}} \right\}\mathcal{N}^{-1},
\end{equation}
where
$\mathcal{N}= \sqrt{{\rho ^2} + {{\left( {{\partial \rho }/{{\partial \psi }}} \right)}^2}}$, which follows from ${\bf{\hat n}} = {{\frac{{d{\bf{\hat t}}}}{{d\psi }}} \mathord{\left/
 {\vphantom {{\frac{{d{\bf{\hat t}}}}{{d\psi }}} {\left| {\frac{{d{\bf{\hat t}}}}{{d\psi }}} \right|}}} \right.
 \kern-\nulldelimiterspace} {\left| {\frac{{d{\bf{\hat t}}}}{{d\psi }}} \right|}}$, where ${\bf{\hat t}} = {{\frac{{d{\bf{x}}}}{{d\psi }}} \mathord{\left/
 {\vphantom {{\frac{{d{\bf{x}}}}{{d\psi }}} {\left| {\frac{{d{\bf{x}}}}{{d\psi }}} \right|}}} \right.
 \kern-\nulldelimiterspace} {\left| {\frac{{d{\bf{x}}}}{{d\psi }}} \right|}}$ is the unit tangent vector. The full equation for the local inclination angle
 \begin{equation}
 \varphi  = \tan^{-1}\left[ {\tan \Phi} \left( {\rho \cos \psi  + \frac{{\partial \rho }}{{\partial \psi }}\sin \psi } \right) \mathcal{N}^{-1} \right].
 \label{eq:localslopeB}
 \end{equation}
 The full expression for the local contact angle is therefore given by substituting \eqn\ref{eq:localslopeB} into \eqn\ref{eq:localcontactanlge}. The projection of the linearized Blake-Haynes (BH) velocity \cite{Blake1969, Blake2006} for contact line motion onto the $x-y$ plane gives the normal velocity of the interface
\begin{equation}
\frac{{\partial {\mathbf{x}}}}{{\partial t}} \cdot {\bf{\hat n}} = {U_0}\cos \varphi \left( {\cos \theta  - \cos {\theta _0}} \right),
\end{equation}
where
\begin{equation}
\frac{{\partial {\mathbf{x}}}}{{\partial t}} = \left\{ {\frac{{dX}}{{dt}} + \frac{{\partial \rho }}{{\partial t}}\cos \psi ,\frac{{\partial \rho }}{{\partial t}}\sin \psi } \right\}.
\label{eq:BH}
\end{equation}
$U_0=\gamma/\eta_{\text{CL}}$ is the contact line's velocity scale defined in terms of the contact line viscosity $\eta_{\text{CL}}$ and the surface tension  of the gas-liquid interface $\gamma$. We assume that the continuous phase (water) only partially wets the solid substrate (silicon nitride) ($\theta_0>0$). Previously, the BH equation has been successfully applied to model the adsorption dynamics of colloids at interfaces \cite{Kaz2012}. Since the observed interfacial velocities in our experiments are well below the threshold required to support a dynamically-formed liquid film \cite{Brochard-Wyart2001}, we assume that the gaseous phase contacts the solid surfaces at all times. In the supplement, we compare bulk and contact line hydrodynamic dissipation with that of BH wetting/de-wetting and determine that the latter dominates the system. Although the disjoining pressure may influence the details of the interface geometry and play a role in wetting and de-wetting dynamics when the gap between the plates is small, this effect is not accounted for by the coarse-grain BH model.

Substititung eqns. \ref{eq:localcontactanlge} and \ref{eq:localslopeB} into \eqn \ref{eq:BH} and linearizing for $\Phi\ll1$ and $\partial\rho/\partial\psi\ll1$ about $\Phi=0$ and $\partial\rho/\partial\psi=0$, we obtain
\begin{multline}
\frac{{\partial \rho }}{{\partial t}} + \frac{{dX}}{{dt}}\left( {\frac{{\sin \psi }}{\rho }\frac{{\partial \rho }}{{\partial \psi }} + \cos \psi } \right) + \mathcal{O}\left(\left( {{{\frac{{\partial\rho }}{{\partial\psi }}}}} \right)^2\right) =\\
U_0\left\{\frac{h_0}{R}+\Phi\left[\frac{X}{R}+\cos\psi\left(\frac{\rho}{R}-H\right)-\frac{\sin\psi H}{\rho}\frac{\partial\rho}{\partial\psi}\right.\right.\\
\left.\left.+\mathcal{O}\left(\left(\frac{\partial\rho}{\partial\psi}\right)^2\right)\right]-\cos\theta_0+\mathcal{O}\left(\Phi^2\right)\right\}
\label{eq:BHunpinned}
\end{multline}
where $H=\sqrt{1-\left(h_0/R\right)^2}$.

In our experiments, we observed bubbles with an advancing front and a pinned aft, \fig\ref{fig:growthsequence}. To accommodate such situations, we modified the BH theory, introducing the weighing pre-factor   $\left(1+\cos\psi\right)/2$ such that \eqn\ref{eq:BHunpinned} becomes

\begin{multline}
\frac{{\partial \rho }}{{\partial t}} + \frac{{dX}}{{dt}}\left( {\frac{{\sin \psi }}{\rho }\frac{{\partial \rho }}{{\partial \psi }} + \cos \psi } \right) =\\
U_0\frac{\left(1+\cos\psi\right)}{2}\left\{\frac{h_0}{R}+\Phi\left[\frac{X}{R}+\cos\psi\left(\frac{\rho}{R}-H\right)\right.\right.\\
\left.\left.-\frac{\sin\psi H}{\rho}\frac{\partial\rho}{\partial\psi}\right]-\cos\theta_0\right\}
\label{eq:BHpinned}
\end{multline}
The above weighting function smoothly reduces the contact line velocity from its maximum value at $\psi=0$ to zero at $\psi=\pi$ .

In defining our coordinate system, we introduced the variable $X$, denoting the position of the bubble's center, without providing a rigorous definition. We do so now by selecting $X\left(t\right)$ such that $\rho\left(0,t\right)=\rho\left(\pi,t\right)$ at all times. Equivalently, $\left.\partial\rho/\partial t \right|_{\psi=\pi}=\left.\partial\rho/\partial t \right|_{\psi=0}$  with $\rho\left(0,0\right)=\rho\left(\pi,0\right)$. Evaluating \eqn\ref{eq:BHunpinned} for $\psi=0$ and $\psi=\pi$, subtracting the results, and incorporating $\left(\partial\rho/\partial t\right)|_{\psi=\pi}=\left(\partial\rho/\partial t\right)|_{\psi=0}$  we obtain the evolution equation for $X$ for the unpinned case

\begin{equation}
2R\left(\frac{dX}{dt}+ {U_0}\Phi H \right)=
{U_0}\Phi \left( {\rho \left( {0,t} \right) + \rho \left( {\pi ,t} \right)} \right)
\label{eq:Xunpinned}
\end{equation}
Following the same procedure but with \eqn\ref{eq:BHpinned} gives the evolution for $X$ for the pinned case 

\begin{multline}
R \left(2\frac{dX}{dt}+U_0 \Phi H \right)=\\
U_0\left\{\left[h_0+\Phi\left(X+\rho\left(0,t\right)\right)\right]-R\cos\theta_0\right\}
\label{eq:Xpinned}
\end{multline}

The evolution of the unpinned contactline is governed by eqns. \ref{eq:BHunpinned} and \ref{eq:Xunpinned}; the pinned contactline is governed by \ref{eq:BHpinned} and \ref{eq:Xpinned}.  To complete our model, we still need to address the dependence of $R$ on time.  For example, in an isobaric process, one would expect $\dot{R}=0$. In the supplement, we consider a Gedanken experiment, wherein the bubble's volume is controlled in a specific way $V\propto t^3$.  Although perhaps not practical, this case is instructive as it admits geometrically self-similar growth in the asymptotic limit $t\rightarrow\infty$, which is helpful for understanding contact line dynamics when the growth rate is independent of $R$. Here, however, we are interested in mass transfer-driven growth, which we address below.

We begin by writing an expression for the volume of the bubble; it can be divided into two contributions, a nearly cylindrical portion
\begin{equation}
{V_\text{I}} =
\int\limits_0^{2\pi } {{\rho ^2}\left( {{h_0} + \rho \cos \psi {\mathop{\rm tan}\nolimits} \Phi } \right){\rm{d}}\psi }
\end{equation}
and the portion that bulges out from the contact line 
\begin{equation}
{V_{\text{II}}} = \int\limits_0^{2\pi } {\frac{{{R^2}}}{2}\left( \begin{array}{l}
\pi  - 2\theta + 2\varphi\\
 - \sin \left( {2\theta - 2\varphi } \right)
\end{array} \right){\cal N}{\rm{d}}\psi }
\label{eq:volII}
\end{equation}

In \eqn\ref{eq:volII}, we assumed $R/\rho\ll1$  to simplify the integrand. The integrals can be further simplified by applying, as before, the linearization, $\Phi\ll 1$ and $\partial\rho /\partial \psi \ll 1$, to yield
\begin{multline}
V=V_{\text{I}}+V_{\text{II}}=\\
\int\limits_0^{2\pi } {\frac{1}{2}\rho \left\{ {2{h_0}\rho  + {R^2}\left[ {\pi  - 2{{{\mathop{\rm cos}\nolimits} }^{ - 1}}\left( {\frac{{{h_0}}}{R}} \right) - \sin \left( {2{{{\mathop{\rm sin}\nolimits} }^{ - 1}}\left( {\frac{{{h_0}}}{R}} \right)} \right)} \right]} \right\}}\\
+ \Phi \left[ \frac{2}{3}{\rho ^3}\cos \psi  - h_0^2\frac{{\partial \rho }}{{\partial \psi }}\sin \psi  + {\rho ^2}\left( {\frac{{2h_0^2\cos \psi }}{{RH}} + X} \right)\right.\\
\left.+ \rho h_0^2\left( {\cos \psi  + \frac{{2X}}{{RH}}} \right) \right]
+ \mathcal{O}\left( {{\Phi ^2}} \right)\mathcal{O}\left( {{{\frac{{\partial \rho }}{{\partial \psi }}}^2}} \right){\rm{d}}\psi
\label{eq:volume}
\end{multline}

Next, we derive the expression for the effective surface area of a bubble available for mass transport. While the bubble is not perfectly cylindrical, the radius of the outward bulge is small compared to the overall radius of the contact line. We therefore expect the concentration field around the bubble to be essentially similar to that around a cylindrical bubble. Hence, the surface area 
\begin{equation}
S\sim\int\limits_0^{2\pi } {\int\limits_{ - h\left( \rho  \right)}^{h\left( \rho  \right)} \mathcal{N} {\rm{d}}z{\rm{d}}\psi }.
\end{equation}
Integration in the $z$-direction gives
\begin{equation}
S = 2\int\limits_0^{2\pi } {\left( {{h_0} + \left( {X + \rho \cos \psi } \right){\mathop{\rm tan}\nolimits} \Phi } \right)\mathcal{N} } {\rm{d}}\psi
\end{equation}
and linearizing yields
\begin{equation}
S=2\int\limits_0^{2\pi } {\rho \left( {{h_0} + \Phi \left( {X + \rho \cos \psi } \right)} \right) + \mathcal{O}\left( {{\Phi ^2}} \right)\mathcal{O}{{\left( {\frac{{\partial \rho }}{{\partial \psi }}} \right)}^2}{\rm{d}}\psi }.
\label{eq:area}
\end{equation}

Next, we estimate the quasi-static total mass flux $\dot{n}$  into a bubble growing in a supersaturated solution with fixed far field concentration $C_{\infty}$ (at $\rho=\rho_{\infty}$) by solving the steady diffusion equation in cylindrical coordinates. We assume that the bubble's contact line is nearly circular with radius $\rho\sim\rho_0$ (where $\rho_0$ is the leading order coefficient describing the shape of the contact line and will be described in \eqn\ref{eq:expansion}). The gas concentration next to the bubble's surface $C_s=C\left(\rho=\rho_0\right)$ is given by a combination of Laplace's equation and Henry's law $C_s=K\left(P_{\infty}+\gamma/R\right)$, where $K$ is the Henry constant. In the above, we again took advantage of $R/\rho\ll1$ by assuming that only the confinement radius $R$ contributes to the Laplace pressure. For simplicity, we assume that the state of the gas inside the bubble is described with the ideal gas equation of state $PV=nBT$, where $P$ is the pressure, $B$ is the universal gas constant, and $T$ is the absolute temperature. Taking the time derivative of the state equation and using eqns. \ref{eq:volume} and \ref{eq:area} gives the differential equation that couples geometry and pressure (radius of curvature)
\begin{multline}
\left( {{P_\infty } + \frac{\gamma }{R}} \right)\dot V - \gamma V\frac{{\dot R}}{{{R^2}}} =\\
 S\frac{{D\left[ {{C_\infty } - K\left( {{P_\infty } + \gamma /R} \right)} \right]}}{{\ln \left( {{\rho _0}/{\rho _\infty }} \right)}}BT
 \label{eq:state}
\end{multline}

We define supersaturation $\alpha$ or the excess dissolved gas concentration relative to the surface concentration of the initial bubble
\begin{equation}
\alpha  = {\log _{10}}\left(C_{\infty}/C_0  - 1 \right),
\label{eq:alpha}
\end{equation}
where ${C_0} = K\left( {{P_\infty } + \gamma /R\left( {t = 0} \right)} \right)$. When supersaturation is large, the system is driven at a rate that is incommensurate with the BH velocity and the contact angle goes to zero somewhere along the contact line. To avoid such a situation, we restrict our analysis to small and moderate supersaturations. 

For concreteness, we consider here circumstances when the initial geometry of the bubble is a spherical section with equilibrium contact angle along the entire contact line. While a spherical bubble does not satisfy our assumption $R/\rho\ll1$, it is the \emph{only} geometry that creates a uniform contact angle along the entire contact line. Any other geometry would introduce additional dynamics at early times, which we wish to avoid. While this initial state is somewhat artificial, we find that, in practice, the condition $R/\rho\ll1$ is reached quickly. \fig\ref{fig:schematic}B depicts the geometry of the initial state of our bubble. Based on simple geometric considerations, we have the following relationship between the initial radius and the height of the channel at the center of the sphere 
\begin{equation}
R_0=R\left(t=0\right)=h_0 \frac{\cos\Phi}{\cos\theta_0},
\label{eq:RIC}
\end{equation}
where $h_0$ is used in the definition of our coordinate system \eqn\ref{eq:localheight2}. The initial contact line shape is a circle with radius $R_0\sin\theta_0$. The projection of this circle onto the $x-y$ plane is an ellipse with major and minor radii $\left\{a,a^* \right\}$ given by
\begin{equation}
\left\{ {\begin{array}{*{20}{c}}
a\\
a^*
\end{array}} \right\} = \left\{ {\begin{array}{*{20}{c}}
{{{R}_0}\sin {\theta _0}}\\
{{{R}_0}\sin {\theta _0}\cos \Phi }
\end{array}} \right\} = \left\{ {\begin{array}{*{20}{c}}
{{h_0}\tan {\theta _0}\cos \Phi }\\
{{h_0}\tan {\theta _0}{{\cos }^2}\Phi }
\end{array}} \right\}
\end{equation}
In our polar coordinate system centered about $X$, the contact line is described by
\begin{equation}
\rho \left(\psi, {t = 0} \right) = \frac{{{h_0}\tan {\theta _0}{{\cos }^2}\Phi }}{{\sqrt {{{\cos }^2}\psi  + {{\cos }^2}\Phi {{\sin }^2}\psi } }}
\label{eq:rhoIC}
\end{equation}
where the initial center of the bubble
\begin{equation}
X\left( {t = 0} \right) =  - {h_0}\sin \Phi \cos \Phi.
\label{eq:XIC}
\end{equation}
As illustrated by \fig\ref{fig:schematic}B, $X\left(t=0\right)$ is negative because the center of the initial contact line is to the left of the center of the initial sphere, which serves as the origin of the global coordinate system. To summarize, the problem statement consists of a partial differential equation (the BH relationship for contact line velocity for unpinned \eqn\ref{eq:BHunpinned} and pinned cases \eqn\ref{eq:BHpinned} ), equation of state \eqn\ref{eq:state}, evolution equation for the bubble's center (unpinned \eqn\ref{eq:Xunpinned} and pinned \eqn\ref{eq:Xpinned}), and initial conditions (eqns. \ref{eq:rhoIC} and \ref{eq:XIC}). 

To simplify the numerical solution of this hybrid system of equations, we assume that $\rho$ is a continuous function of $\psi$ and use a spectral decomposition of the contact line's position
\begin{equation}
\rho \left( {\psi ,t} \right) = \sum\limits_{n = 0}^N {{\rho _n}\left( t \right)\cos \left( {n\psi } \right)}.
\label{eq:expansion}
\end{equation}
Only cosine terms are used in the above expansion because we restrict our analysis to bubbles symmetric with respect to the principal axis of the wedge  $\psi=0,\pi$). We substitute \eqn\ref{eq:expansion} into the BH equation (eqns. \ref{eq:BHunpinned} or \ref{eq:BHpinned})  and require that it is satisfied in the sense of the weighted residuals
\begin{equation}
\int\limits_0^{2\pi } {\cos \left( {n\psi } \right)\left( {\text{BH}_{\text{LHS}} - \text{BH}_{\text{RHS}}} \right){\rm{d}}} \psi  = 0,{\rm{   }}n = 0,1,2 \cdots.
\end{equation}
We show terms for the case $N=2$ in the electronic supplement. The above decomposition has the virtue of not only reducing the system to a set of non-linear ordinary differential equations, but also readily incorporating integral (\eqn\ref{eq:state}) and evolution equations (\eqn\ref{eq:Xunpinned} or \eqn\ref{eq:Xpinned})  in the model. We find rapid convergence and very few modes are needed when no-pinning is used (\emph{i.e.} when bubbles are free to translate, the contact line is nearly circular at all times). In the presence of pinning, the model with $N=$5 functions well for moderate time and is able to model bubble growth to the point they detached in the experiment. For long times, higher frequency modes begin to dominate, which invalidates assumptions made in our model. While we only use cosine terms, the problem could be generalized to include sine terms if, for example, initial conditions with broken symmetry were of interest or if one wished to include additional forces acting transverse to the wedge axis; otherwise, there is no physical reason for information to flow into the odd functions given the problem's symmetry.

\section{Experimental Method and Observations}

We imaged bubble nucleation, growth, and migration in our custom-made liquid cell (the nanoaquarium) \cite{Grogan2010,Grogan:2011dj, Grogan2014} with a transmission electron microscope (TEM). The liquid cell consists of a thin (nominally 200 nm) liquid layer confined and hermetically sealed between two very thin (50nm) silicon nitride membranes (100 $\mu$m x 100 $\mu$m). See \fig S1 in the electronic supplement for a schematic depiction of our liquid cell. Additional details on the structure of the liquid cell and its fabrication are available in Grogan and Bau \cite{Grogan2010}. The liquid is sealed from the vacuum of the electron microscope and the entire assembly is thin enough to be transparent to electrons.  In our experiments, the liquid cell is filled with a solution of water with trace amount of the surfactant cetrimonium bromide (CTAB). The liquid cell is imaged at 300kV using a beam of diameter 4 $\mu$m and current 1nA. \fig\ref{fig:growthsequence} (movie S1) shows a series of bubbles nucleating from a location that is most likely a defect such as a pit on the interior surface of one of the liquid cell windows. These bubbles form because of radiolysis of water by the electron beam that generates gaseous species. The primary species are hydrogen and oxygen \cite{Schneider2014}, their production induces bubble nucleation and sustains bubble growth. Liquid cell electron microscopy is a new that can provide nanometer resolution of aqueous samples \cite{Ross2015}. In terms of imaging fluid dynamics phenomena, the technique is still evolving: the liquid layer geometry can not be defined accurately by the microscopist and the nucleation sites for bubbles are determined by random defects in the silicon nitride windows. This restricts our observations to somewhat uncontrolled experiments and precludes quantitative comparison between our theoretical predictions and data.  Nevertheless, the liquid cell provides sufficient information to allow us to qualitatively compare our theoretical predictions with experimental data. TEM liquid cell microscopy has been used previously to image condensation \cite{Bhattacharya:2014cr}, motion of droplets \cite{Mirsaidov2012a}, dewetting phenomena \cite{Mirsaidov2012b} and bubble formation by heating \cite{White2012} however, the physics behind these phenomena have not been modeled extensively and appear distinct from the phenomena discussed here.

\begin{figure}[h]
\includegraphics[width=\columnwidth]{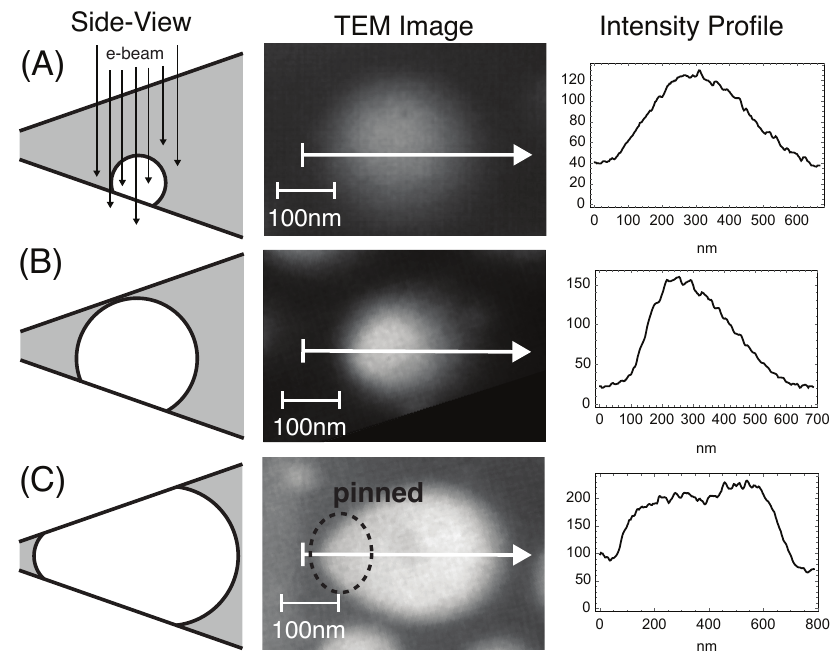}
\caption{(A-C) A series of schematics (1st column), bright field electron micrographs (2nd column) and intensity profiles taken alone the midline of the bubble indicated on the images (3rd column) of a bubbles growing in under confinement. The bubble in (C) is pinned to a nucleation site.}\label{fig:growthstages}
\end{figure}

\begin{figure}
\includegraphics[width=3in]{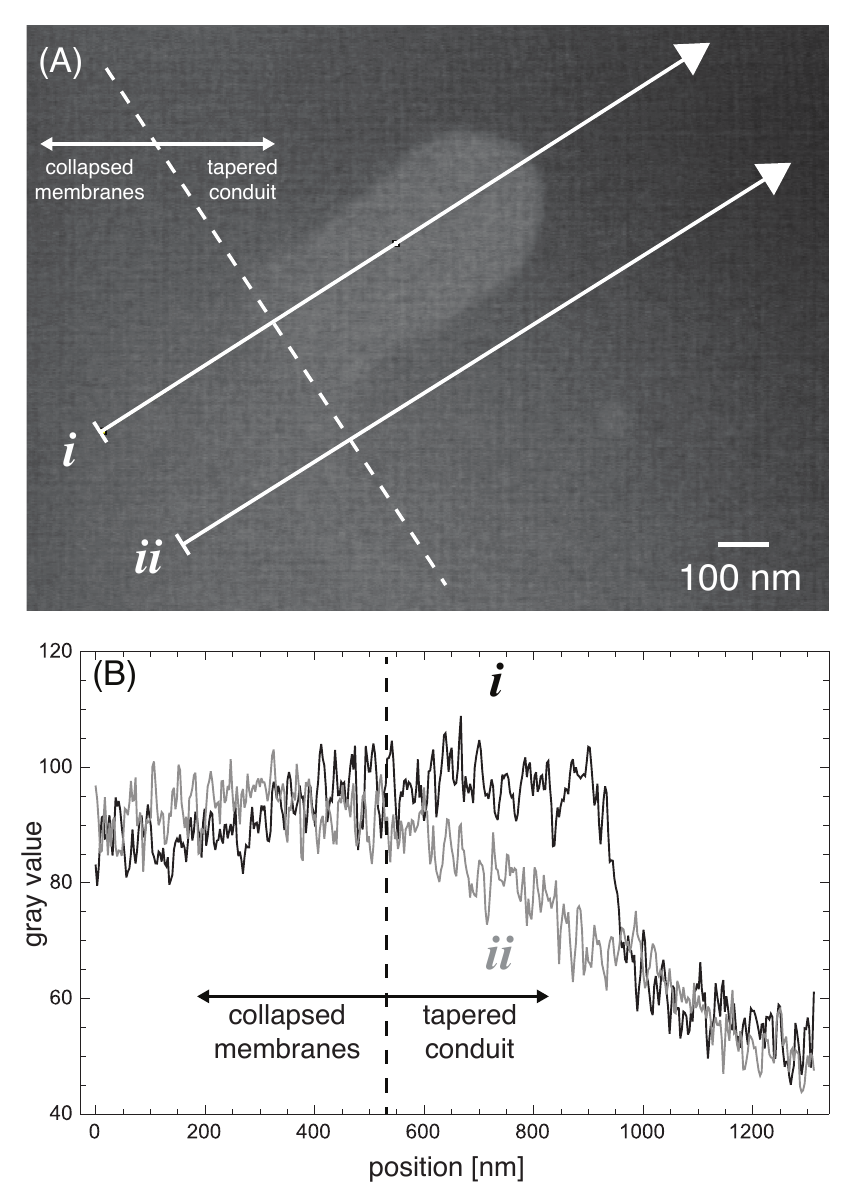}
\caption{(A) and (B), respectively, show a microscopy image and corresponding intensity profiles ( \emph{i} is taken along the centerline of the bubble, while \emph{ii} is taken parallel to it in a nearby region of the device ). Since the transmission of (\emph{ii}) is flat and similar to that of the bubble (\emph{i}) we conclude that the membranes have contacted, along the line labled in (A) and (B).}
\label{fig:collapsenano}
\end{figure}

\fig\ref{fig:growthstages}A-C shows bright field electron microscope images of bubbles growing under confinement in our liquid cell. Three bubbles at different stages in their growth are shown along with a schematic side view (first column) that interprets the intensity profiles (third column). The scatter of electrons by the irradiated medium and the darkness of the image scale with the integrated atomic number density along the beam's path, or in our case, the local thickness of water. Darker (lower intensity) sections of the image therefore indicate a thicker liquid layer. The flat intensity profile of the bubble \fig\ref{fig:growthstages}C indicates that the bubble has contacted both membranes, while the ``shaded'' bubbles in (A) and (B) indicate that electrons encounter both liquid and vapor along their path and that the bubbles have therefore not contacted both surfaces.

When we observe nucleation, the bubbles appear to emanate periodically from a presumed impurity in the silicon nitride film. \fig\ref{fig:growthsequence} shows snapshots of bright field electron microscope micrographs of a bubble growing under high confinement (movie S1). We compare our theory to bubbles like these. Nucleated bubbles are first observed when they are $\sim80$ nm in diameter and depart when they are $\sim250$nm in size. The bubbles depart by breaking into two bubbles: a smaller bubble that remains attached to the nucleation site and a larger bubble that continues to translate. The process repeats with a frequency of 0.3 Hz.  The reproducibility of the process from one bubble to the next suggests quasi-static far field supersaturation \cite{Grogan2014}. Indeed, theory suggests that the radiolysis process is self-regulating and that the concentration of radiolysis products achieve steady state \cite{Schneider2014}. Since we cannot measure the dissolved gas concentration, we use it as a fitting parameter in our model, $\alpha$ \eqn\ref{eq:alpha}. 

The experimental data suggests that locally, bubbles always migrate in the same direction towards lower confinement, \fig\ref{fig:growthsequence}B.  In support of this, we show a bubble at a different nucleation site \fig\ref{fig:collapsenano}A; \fig\ref{fig:collapsenano}B depicts two intensity profiles along parallel lines (\emph{i}) and (\emph{ii}), shown as arrows in \fig\ref{fig:collapsenano}A.  Profile (\emph{i}) includes the bubble and profile \emph{ii} goes alongside the bubble.  The recorded intensity declines as we go from left to right along the line (\emph{ii}), suggesting that the thickness of the liquid layer increases from left to right. Furthermore, the similar transmitted intensity of the bubble compared to the region from which it emanates indicate a very thin liquid layer and possibly membrane contact. During loading of the liquid cell, capillary forces can pull the silicon nitride membranes into close proximity. We show such collapsed membranes in the supplement at lower magnification using optical microscopy (\fig S2). \fig\ref{fig:collapsenano} and \fig S3 show that bubbles grow and migrate in the tapered conduit in the direction of diverging plates.  This is the situation for which we built our model in Section II above.  Like \fig\ref{fig:growthstages}C, the detected intensity along the bubble (\emph{i}) is nearly uniform and the transition from the bubble to the bulk liquid is relatively rapid compared to the size of the bubble's plateau, one can infer that the bubble is highly confined, justifying the assumption made in the theory section $R/\rho \ll 1$. 

Using simple image processing algorithms, we automate measurements of bubble features such as centroid position, area, and contact line shape and stitch together bubble trajectories using established techniques \cite{Schneider:2016hm,Blair}. In the Section V, we compare experimental observations with theoretical predictions.

\section{Model Results and Discussion}

\begin{figure}[h]
\includegraphics[width=\columnwidth]{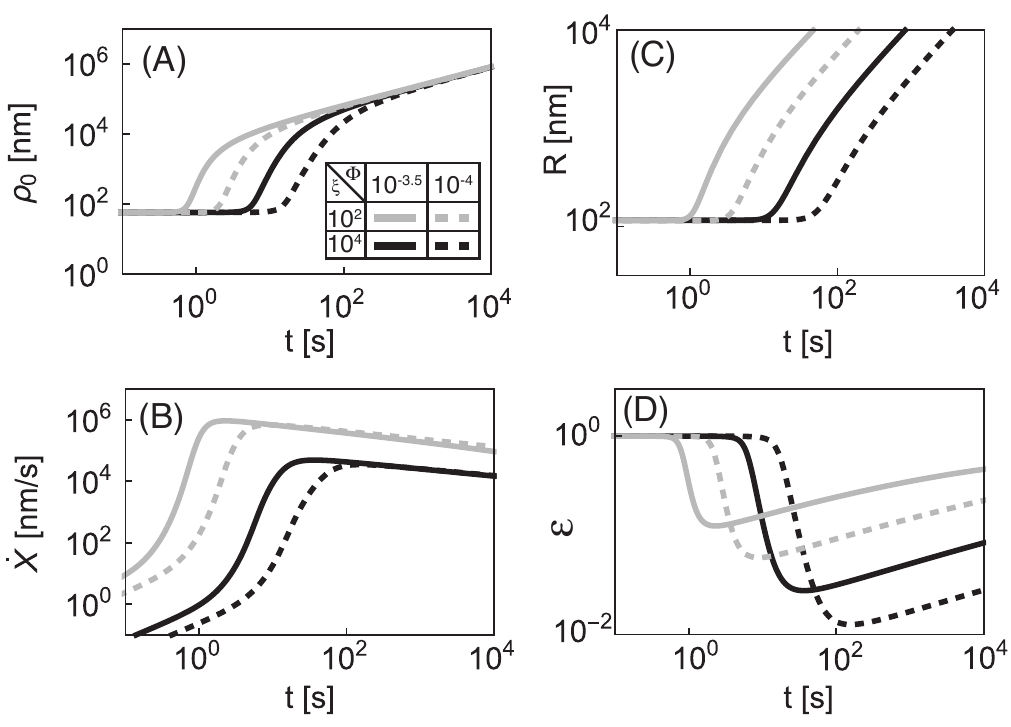}
\caption{Model predictions for (A) bubble size, (B) velocity and (C) radius of curvature and (D) aspect ratio for two different slopes and contact line viscosities:  $\xi=10^2$ (gray) and $\xi=10^4$(black), and slopes $\Phi=10^{-4}$ (dotted) and $\Phi=10^{-3.5}$ (solid) . In all cases $h_0=100$ nm and $\alpha=-6$. \label{fig:theory}}
\end{figure}

\begin{figure}[h]
\includegraphics[width=\columnwidth]{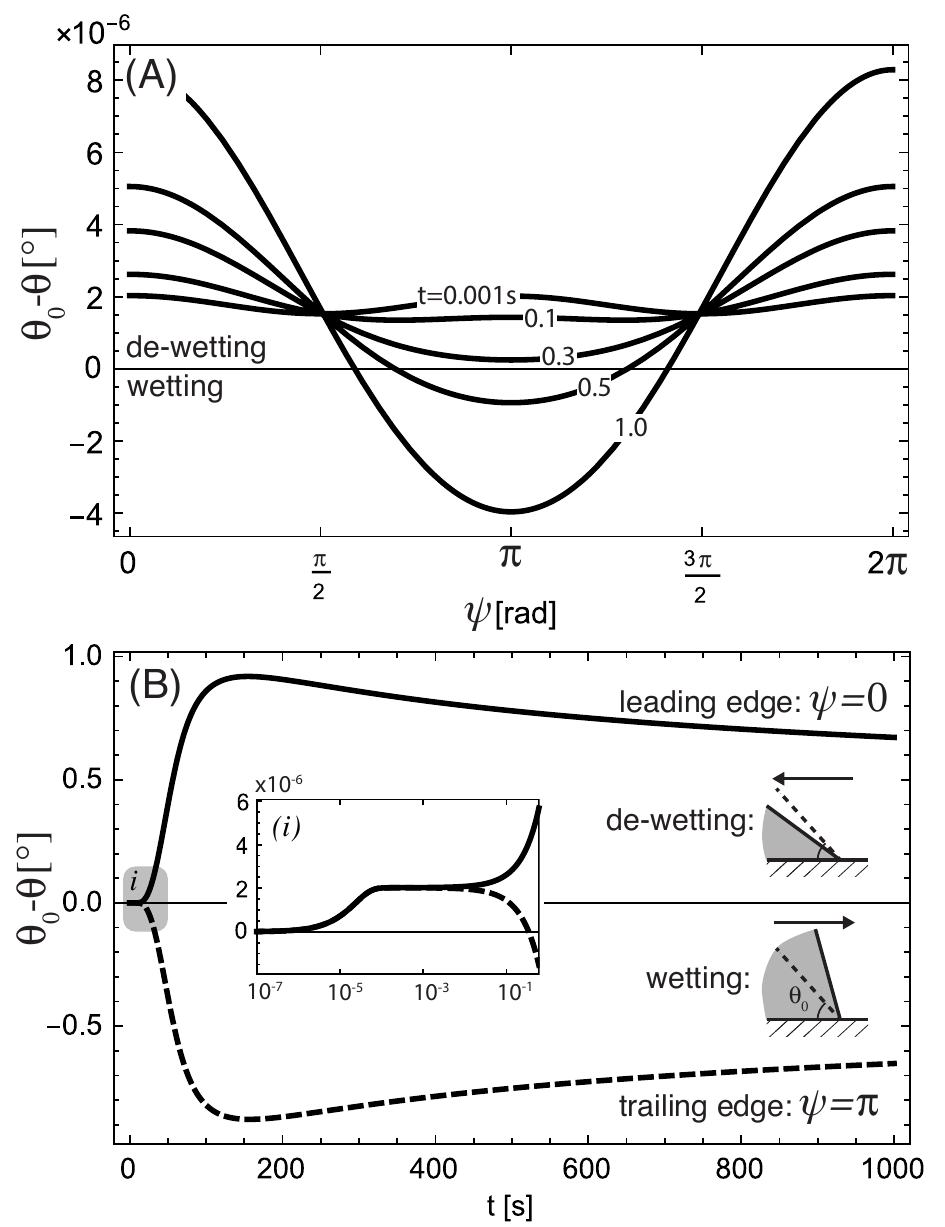}
\caption{(A) Contact angle $\theta\left(\psi,t\right)$ relative to the equilibrium contact angle $\theta_0$ as a function of polar angle $\psi$ at a few different stages in bubble growth. (B) $\theta_0-\theta$ at the leading (solid) and trailing (dashed) edges of the bubble as a function of time; for $t>0.3$ s dewetting occurs at the leading edge and wetting occurs at the trailing edge. Inset (\emph{i}) shows short dynamics in which, the bubble expands in all directions. $h_0=100$nm, $\alpha=-6$, $\xi=10^4$, and $\Phi=10^{-4}$. \label{fig:contactangle}}
\end{figure}

In all our theoretical predictions, we assume that the gaseous species is hydrogen and that the liquid cell is at room temperature ( $T=298$ K) and pressure ( $P=0.1$ MPa) \cite{Grogan2014}. We assume that as bubbles grow they do not significantly increase the pressure of the device since their volume is miniscule to that of the device.  We use  $\theta_0=30^{\circ}$ \cite{Arkles2006}, $K_{\text{H}_2}=7.74\times10^{-6}$ mol/Pa $\text{m}^3$ \cite{Sander2015},  $D_{\text{H}_2}$ $\text{m}^2$/s \cite{Cussler2009},  $\eta_{\text{CL}}=k_{\text{B}}T/\left(\kappa_0\lambda^3\right)=\xi\eta_0$ Pa s  \cite{Petrov1997,Blake2006}, bulk viscosity $\eta_0=8.9\times 10^{-4}$ Pa s, and surface tension of the vapor-liquid interface and $\gamma=40$ mN/m. We use surface tension of the gas-liquid interface lower than that of pure water to account for the presence of trace amounts of the CTAB surfactant in our experiment \cite{Berg2010}. The negative value for $\alpha$ corresponds to a far field gas concentration that is only slightly above the equilibrium concentration at the surface of the initial bubble.  In the expression for the contact line viscosity $\lambda$ is the lattice spacing, $k_\text{B}$ is the Boltzman constant, and  $\kappa_0$ is the hopping frequency. We explore a range of  dimensionless contact line viscosities ($\xi=\eta_{\text{CL}}/\eta_{0}=10^2-10^6$) because literature values vary greatly ($\lambda\sim10^{-10}-10^{-9}$ m and  $\kappa_0\sim10^3-10^9$ $\text{s}^{-1}$  \cite{Blake2006}), are not known for our system, and may be altered by the presence of surfactants \cite{Petrov1997}. The local slope of the device is also not precisely known, but estimated to be in the range $\Phi=10^{-3}-10^{-2}$ as further discussed in the  electronic supplement. 

\begin{figure}[h]
\includegraphics[width=\columnwidth]{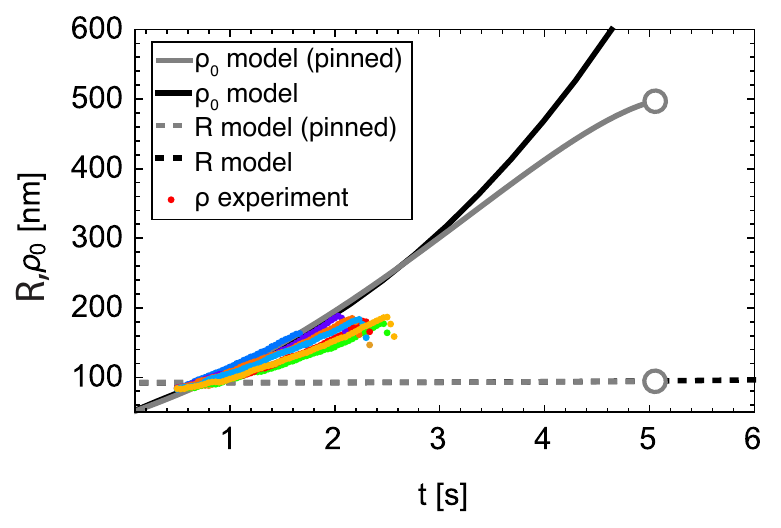}
\caption{Comparison of measured (points) and predicted average bubble size $\rho_0$ as a function of time for several bubbles emerging from the same nucleation site (colors indicate different bubbles nucleating from the same site). Model results for unpinned (black) and pinned (gray) are shown along with the evolution of the radius of curvature $R$ (dashed). The open circle denotes the end of the validity of the pinned model. \label{fig:comparemetrics}}
\end{figure}

Before comparing to experiment, we examine the impact of some of the key parameters in the model: the slope of the channel $\Phi$ and contact line viscosity $\xi$ in \fig\ref{fig:theory}. We fix initial channel height $h_0=100$nm and supersaturation $\alpha=-4$. \fig\ref{fig:theory} depicts (A) the bubble size $\rho_0$, (B) the velocity of the bubble's center of mass $\dot{X}$, (C) the radius of curvature R, and (D) the ratio of curvatures  $R/\left(\rho_0+R\left(1-\sin\theta_0\right)\right)$ as functions of time when $\xi=10^2$ (gray line) and $10^4$ (black line) and when $\Phi=10^{-3.5}$ (solid line) and $10^{-4}$ (dashed line). The effect of the contact line viscosity is intuitive. The greater the dissipation at the contact line, the more the contact angle must be pushed out of equilibrium to achieve the same velocity leading to slower growth. Interestingly, changing the slope of the taper also changes the time at which the bubble will appreciably depart from its initial size. To understand why this is the case we consider the short time dynamics more closely.

Initially, the bubble grows slowly because it takes time for the contact line to move. Contact-line motion is needed to enable the Laplace pressure and the equilibrium concentration of the dissolved gas next to the bubble's surface to decrease. At short times, before the contact line moves, the only way to accommodate mass is by \emph{decreasing} $R$, which bulges the bubble further beyond the contact line, increasing Laplace pressure, slowing mass transfer. Thus at short times, contact line resistance introduces a negative feedback mechanism on bubble growth. This dynamic is apparent when examining the contact angle distribution at short times ($t<0.3$ s) in \fig\ref{fig:contactangle}, the local contact angle $\theta\left(\psi,t\right)$ is for all $\psi$ less than the equilibrium contact angle $\theta_0$ indicating that the bubble has displaced fluid in all directions. 

According to BH theory, such a contact angle distribution will causes the contact line to advance into the liquid (de-wet) and give way to the growing bubble. While this occurs at all points on the contact line initially, the contact angle distribution eventually breaks symmetry giving rising to a faster velocity at the leading edge, \fig\ref{fig:contactangle}. As the interface at the leading edge moves $h$ increases, allowing $R$ to increase, the equilibrium concentration of the dissolved gas to decrease, and the mass transport to increase. In other words, the increasing channel height provides a positive feedback mechanism that accelerates the bubble's growth once the contact line begins moving. The larger the opening slope or smaller the contact line viscosity, the quicker the switch between negative and positive feedback occurs. As the bubble's geometry evolves, the contact angle distribution switches from being non-uniformly below the equilibrium contact angle $\theta_0$ (corresponding to de-wetting) to a portion of the rear contact line being greater than the equilibrium value. In other words, once the bubble begins translating down the conduit, de-wetting occurs at the leading edge ($\psi=0$), and wetting occurs at the trailing edge ($\psi=\pi$), \fig\ref{fig:contactangle} tracks the evolution of the contact angle at these two locations.  The moment of the sign change of the trailing contact line is made clear in the inset of \fig\ref{fig:contactangle}B\emph{i} (dashed lined).

Interestingly, while the velocity $\dot{X}$ initially increases (\fig\ref{fig:theory}B), it achieves a maximum velocity and then slowly declines as $t\rightarrow\infty$. This velocity scaling is correlated with the aspect ratio $\varepsilon=R/\left(\rho_0+R\left(1-\sin\theta_0\right) \right)$ tending towards unity, indicating that the bubble's geometry is becoming more spherical over time. Since this must also means that the contact angle distribution is becoming more uniform (tending towards $\theta\left(\psi\right)\rightarrow \theta_0$), we consider the limit of fast contact line relaxation $\eta_{\text{CL}}=0$ in the supplement. In this case, the  contact angle is always uniform and at equilibrium; the bubble is therefore a spherical section that grows self-similarly and translates down the conduit at a velocity required to satisfy geometric constraints. In the supplement we show that $\dot{X}\propto t^{-1/2}$, just like \fig\ref{fig:theory}B for large times. For the parameters used, our theory therefore predicts that confinement and contact line dynamics become less important once bubbles are on the order of $\sim$ 100 $\mu$m - 1 mm in extent.

\begin{figure*}
\includegraphics[width=\textwidth]{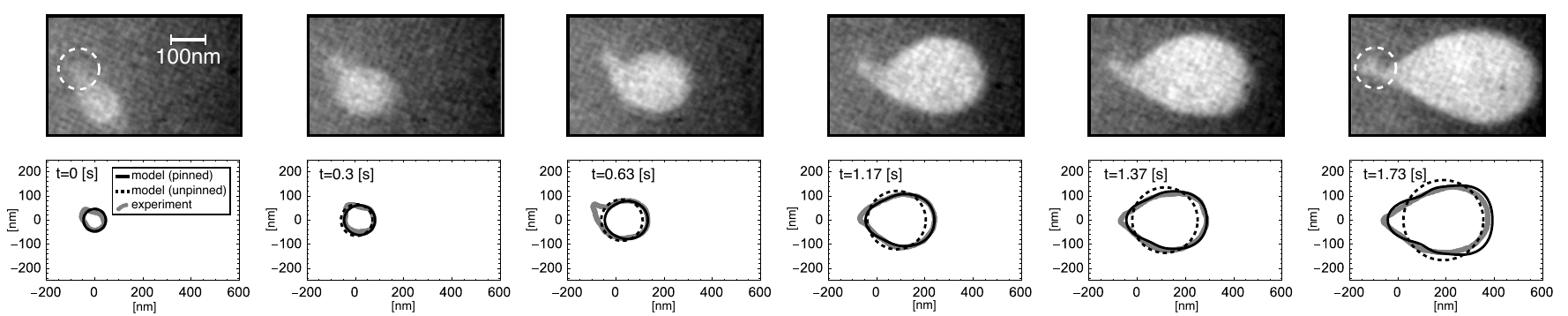}
\caption{: (Top) Experimental observations of a bubble growing from a nucleation site (dashed white lines). (Bottom) Pinned (black lines, \eqn\ref{eq:BHpinned}) and unpinned (dashed lines, \eqn\ref{eq:BHunpinned}) model predictions compared to contact line from top row (thick gray lines).\label{fig:compareshapes}}
\end{figure*}

\section{Comparison with Experimental Observations}

\fig\ref{fig:comparemetrics} compares our model predictions, with (gray) and without (black) pinning (modeled with the weighted \eqn\ref{eq:BHpinned} that smoothly interpolate between a completely pinned contact line at $\psi=\pi$ and full mobility at $\psi=0$), to experimental observations of bubbles that are pinned to their nucleation site (symbols). While we know the bubbles to be pinned, for comparison we show predictions made by both pinned and unpinned versions of the model. For both models,  $\Phi=10^{-3}$, $h_0=80$nm, $\alpha=-2.9$, and $\xi=5.4$. \fig\ref{fig:comparemetrics}A depicts  $\rho_0$ (solid line) and $R$ (dashed line) as functions of time. When $t<10$s, we find no significant difference in the leading order growth rates predicted in the absence and presence of pinning. This is because the feedback of higher order geometric terms  $\rho_{1,2,\cdots}$ on $R$ is weak when the bubble nearly circular. The open circle in \fig\ref{fig:comparemetrics} indicates the end of theoretical predictions for the pinned case; while the model fails at finite time, we note that in the experiment the bubble breaks apart prior to this point. In this paper, we do not consider instabilities that may lead to detachment. Before breakup, there is good qualitative agreement between the model and theory. In \fig\ref{fig:compareshapes}, we more closely compare the predicted geometry of the contact line to observations of the same data set.  We see that the theoretical predictions are in good agreement with the observations away from the pinning/nucleation site (dashed circle).

\begin{figure}[h]
\includegraphics[width=\columnwidth]{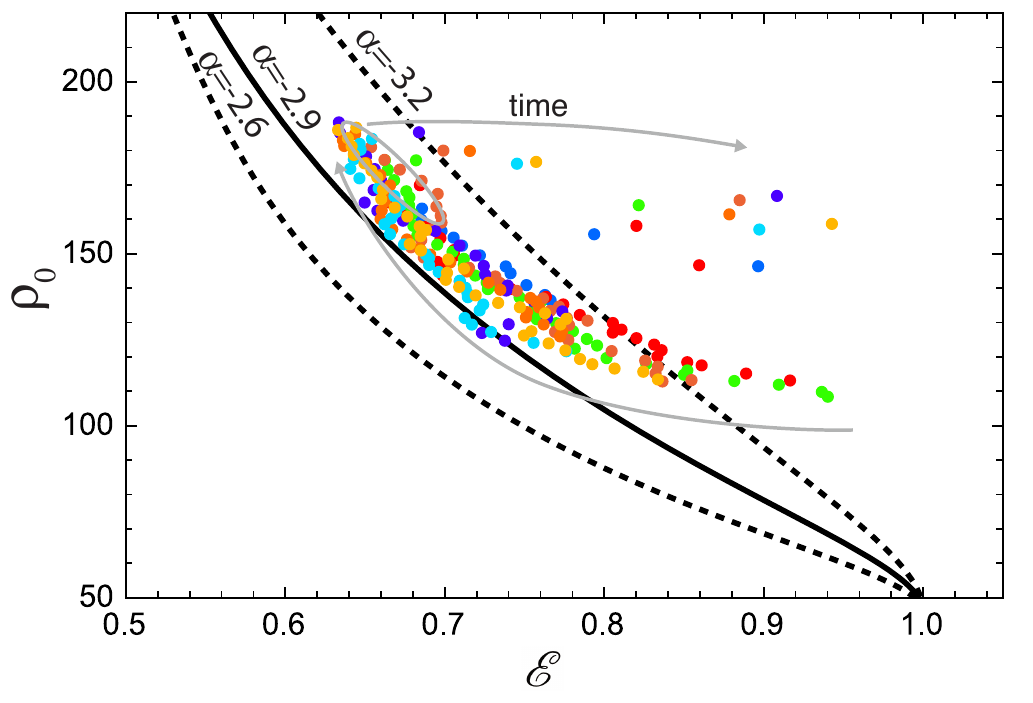}
\caption{Model predictions (lines) of bubble size $\rho_0$ as a function of aspect ratio $\mathscr{E}$ (\eqn\ref{eq:aspectratio}) for three different supersaturation values compared to experiment (points, colors indicate different bubbles nucleating from the same site). Arrows show the progression of time and circle highlights the point at which the bubbles break away from their nucleation sites ($t\sim$1-2 s).}
\label{fig:aspectratio}
\end{figure}

Finally, in \fig\ref{fig:aspectratio} we examine the relationship between the size $\rho_0$ and the \emph{projected} aspect ratio $\mathscr{E}$ of the bubble (not to be confused the aspect ratio $\varepsilon$ plotted in \fig\ref{fig:comparemetrics}D)
\begin{equation}
\mathscr{E}=\left(\rho_0-\rho_2+\rho_4 \right)/\left(\rho_0+\rho_2+\rho_4\right)
\label{eq:aspectratio}
\end{equation}
where $\rho_{2,4}$ are shape corrections to $\rho_0$ (for $N=$5).  We find that the faster growing bubbles (those with higher superstation) become elongated at smaller size. Our model therefore predicts that contact line dynamics and the tapered channel together introduce a growth rate-dependent geometric evolution, which would otherwise be absent in the zero capillary and Bond number regime we consider.  Since bubble breakup likely depends on the extent to which the bubbles elongate, we posit that bubble creation frequency and departure size will depend on super-saturation. Our current single-curvature model does not allow us model the breakup mechanism. Before breakup, a neck forms which invalidates assumptions of our model; however, controlled experiments on droplet breakup in tapered conduits in the quasi-static regime strongly implicate a geometric instability\cite{Dangla2013}.

\section{Conclusion}



We imaged and analyzed the growth and motion of sub-micron bubbles in a supersaturated liquid confined between two narrowly separated, rigid, diverging plates in the asymptotic limit of zero Capillary and Bond numbers. To enable imaging sub-micrometer-size bubbles, we used the experimental technique liquid cell electron microscopy to  provide spatial and temporal resolution that is not readily available by any other types of microscopy. The motion of macroscopic bubbles is often driven by capillary and buoyancy forces, both of which are negligible at the length scales considered here.  We hypothesize that bubble motion and growth are rate-limited by contact line dynamics. To predict contact line motion, bubble growth, and interface geometry, we use the Blake-Haynes mechanism to describe contact line motion as a function of contact angle.  Variations in contact angle result from gas mass flow into the bubble under supersaturated conditions. Our theoretical predictions agree qualitatively with experimental observations.  At early stages, bubbles grow slowly as contact line-mediated curvature mitigates the positive feedback mechanism that would otherwise enhance mass transfer into the bubble and drive rapid growth. This is somewhat similar to one of the more recent mechanisms that has been proposed to explain the longevity of surface-bound nano-bubbles; the pinning of the contact line controls curvature \cite{Weijs2013,Lohse2015a}. At longer times, our model predicts that the growth rate of our bubbles accelerates due to positive feedback: decreased radius of curvature reduces the equilibrium gas concentration next to the bubbles' surfaces, enhancing mass transport, and decreased confinement. 

Our observations and model predictions have implications for various processes where bubbles nucleate and grow within surface defects, fissures, or tailored nanostructures such as are used for boiling and hydrolysis. Clearly, one could exploit geometry to clear bubbles; this much has been explored for microgravity applications \cite{Jenson2014},  albeit relying on driving forces that are negligible at the nanoscale as in our experiments.

Perhaps more intriguing is our model's implication that bubble geometry depends on growth rate. This is somewhat unexpected given the overdamped regime but arises as the result of an additional time scale in the problem, the contact line relaxation. Growth-rate dependence is particularly apparent when aft contact line pinning is included in the model. Our prediction that the aspect ratio of growing teardrop bubbles can be controlled by controlling the supersaturation level is a novel mechanism that could be exploited in device design.

This paper has examined a fluid mechanical problem with the emerging experimental method of liquid cell electron microscopy.  We demonstrate that liquid cell electron microscopy with its few nanometer and video-rate resolution can provide meaningful data that would be difficult, if not impossible, to obtain by other means. Although our liquid cell  imaging provided qualitative information to support our theoretical predictions, the method is still at its infancy  in terms of yielding quantitative data. Some calibration tools exist, for example the use of electron energy loss spectroscopy to measure liquid thickness \cite{Jungjohann2012}, but the accuracy is not sufficient for these types of experiments.   We hope that in the future, liquid cells will be better equipped with diagnostic tools such as means to measure sufficiently accurately the distance between the confining plates and the conditions (pressure, temperature, chemical composition) of the liquid inside the liquid cell.  Such developments would allow for better controlled experiments and will enable quantitative comparisons between theoretical predictions and experimental data.

\section{Acknowledgements}
The research was supported, in part, by the National Science Foundation CBET 1066573 and the Nano/Bio Interface Center through the National Science Foundation NSEC DMR08-32802. All movies were recorded by Joseph. M. Grogan at the IBM T. J. Watson Research Center.



\end{document}